\documentclass{article}
\usepackage{graphicx}
\graphicspath{%
    {converted_graphics/}
    {/}
}
\begin{document}

\begin{center}
{\LARGE Toward a Comprehensive Model of Snow Crystal Growth:}\vskip6pt

{\LARGE \ 4. Measurements of Diffusion-limited Growth at -15 C}\vskip16pt

{\Large Kenneth Libbrecht}$^{1}${\Large , Christopher Miller}$^{2}${\Large ,
Ryan Potter}$^{3}${\Large ,}\vskip4pt

{\Large Nina Budaeva}$^{1}${\Large , Cameron Lemon}$^{3}${\Large , Sarah
Thomas}$^{3}${\Large \ }\vskip8pt

$^{1}${\large Department of Physics, California Institute of Technology}%
\vskip1pt

{\large Pasadena, California 91125}\vskip4pt

$^{2}${\large Department of Physics, University of North Carolina at Chapel
Hill}\vskip1pt

{\large Chapel Hill, North Carolina 27599}\vskip4pt

$^{3}${\large Department of Physics, University of Cambridge}\vskip1pt

{\large Cambridge, England CB3 0HE} \vskip3pt

(send correspondence to \it{kgl@caltech.edu})

\vskip18pt

\hrule\vskip1pt \hrule\vskip14pt
\end{center}

\textbf{Abstract.} We present measurements of the diffusion-limited growth
of ice crystals from water vapor at different supersaturation levels in air
at a temperature of -15 C. Starting with thin, c-axis ice needle crystals,
the subsequent growth morphologies ranged from blocky structures on the
needle tips (at low supersaturation) to thin faceted plates on the needle
tips (at high supersaturation). We successfully modeled the experimental
data, reproducing both growth rates and growth morphologies, using a
cellular-automata method that yields faceted crystalline structures in
diffusion-limited growth. From this quantitative analysis of well-controlled
experimental measurements, we were able to extract information about the
attachment coefficients governing ice growth under different circumstances.
The results strongly support previous work indicating that the attachment
coefficient on the prism surface is a function of the width of the prism
facet. Including this behavior, we created a comprehensive model at -15 C
that explains all the experimental data. To our knowledge, this is the first
demonstration of a kinetic model that reproduces a range of
diffusion-limited ice growth behaviors as a function of supersaturation.

\section{Introduction}

Our overarching goal in this series of investigations is to develop a
comprehensive model of ice crystal growth from water vapor, one that can
reproduce quantitative growth rates as well as growth morphologies over a
broad range of circumstances. Although ice crystal formation has been
studied extensively for many decades, our understanding of the physical
effects governing growth behaviors at different temperatures and
supersaturations remains rather poor \cite{nakaya54, mason63, lamb72,
kurodalac82, lacmann83, nelsonknight98, libbrechtreview05}. For example, the
complex dependence of ice growth morphology on temperature, exhibiting
several transitions between plate-like and columnar structures \cite%
{morph04, hallet09}, remains essentially unexplained even at a qualitative
level, although it was first reported over 75 years ago \cite{nakaya54}.

To address this problem, we have undertaken an experimental program designed
to create small ice crystals with simple morphologies and measure their
subsequent growth under carefully controlled conditions, to an extent and
accuracy surpassing previous efforts \cite{knight12, hallet12, maruyama05,
nelson01, kuroda84}. We model the experimental data using a recently
developed cellular-automata numerical method that can generate physically
realistic faceted structures in diffusion-limited growth \cite{reiter05,
gg08, gg09, kglca13, kglSDmodel, kelly13}. The comparison between measured
and modeled ice crystals then provides valuable information about the
attachment kinetics governing ice growth from water vapor. From this
information we hope to develop a detailed physical picture of the molecular
structure and dynamics of the ice surface during solidification.

\section{Ice Growth Measurements in a Dual Diffusion Chamber}

The ice growth measurements described in this paper were obtained used the
dual diffusion chamber described in \cite{kgldual14}. The first of the two
diffusion chambers was operated with a high water-vapor supersaturation in
air, and in this chamber we grew electrically enhanced ice needles with tip
radii \symbol{126}100 nm and overall lengths of \symbol{126}3 mm, with the
needle axis along the c-axis of the ice crystal. The needle crystals were
then transported to the second diffusion chamber, where the temperature and
supersaturation were independently controlled, and the subsequent growth was
recorded using optical microscopy. A linear temperature gradient in the
second chamber ensured that convection currents were suppressed and that the
supersaturation could be accurately modeled.

\begin{figure}[t] 
  \centering
  \includegraphics[width=4.8in,keepaspectratio]{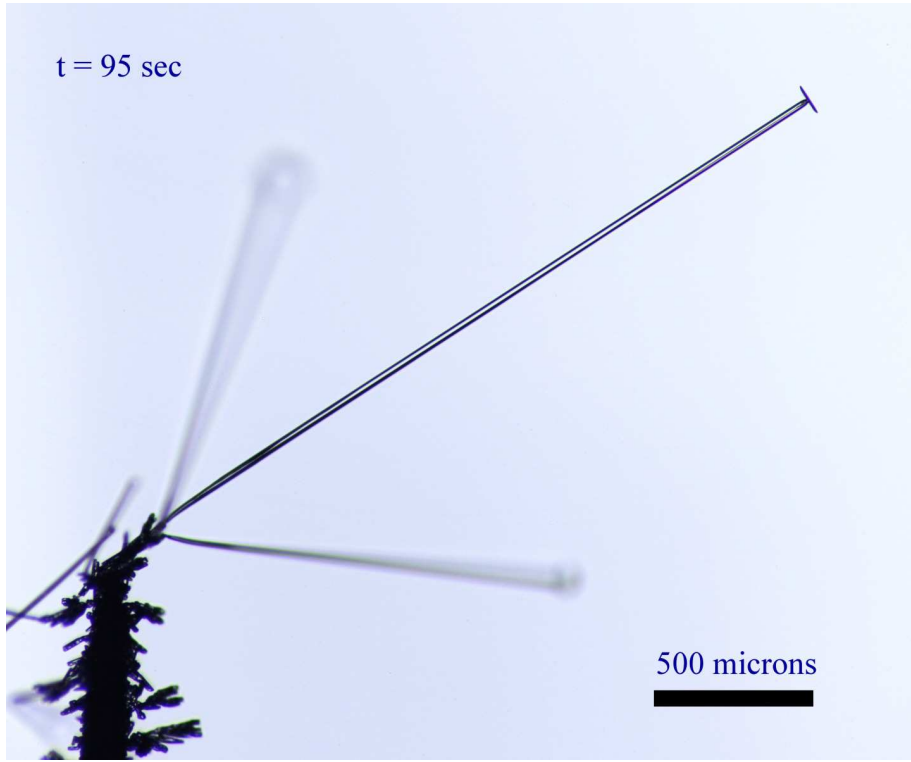}
  \caption{This photograph shows a typical
ice needle growth measurement at -15 C. The tip of the wire substrate is
seen in the lower left of the photo, covered with a large number of frost
crystals. Several thin, c-axis ice needles grew out from the wire tip, and
one was brought into focus with the entire needle in the image plane. After
95 seconds of growth at a water-vapor supersaturation of $\protect\sigma %
_{center}\approx 11$ percent (in this particular example) a thin ice plate
can be seen growing on the tip of the ice needle, here seen in side view.
The ice plate diameter, ice needle diameter below the plate, and the overall
needle length could be extracted from the calibrated image.}
  \label{example}
\end{figure}

Immediately after an ice needle assembly was moved to the second diffusion
chamber, the wire base holding the needles was rotated so a particular test
needle was in focus in the microscope with the needle entirely in the focal
plane, providing a side view of the subsequent ice growth, as shown in
Figure \ref{example}. During this transport and focusing step, a thin,
frost-covered, horizontal shutter plate was positioned just above the ice
needles, reducing the supersaturation below the plate to near zero. Once the
test needle was satisfactorily positioned (typically taking 10-20 seconds),
the shutter was removed and growth measurements commenced. The
supersaturation near the test crystal relaxed to steady state in a time of
order $\tau \approx L^{2}/D\approx 5$ seconds, where $L\approx 1$ cm is the
shutter size and $D\approx 2\times 10^{-5}$ m$^{2}/$sec is the diffusion
constant for water vapor in air.

For all the measurements described below, the air temperature surrounding
the ice needles was maintained at $T=-15\pm 0.1$ C, as determined by a small
calibrated thermistor that was periodically placed at the center of the
diffusion chamber, at the same location as the ice needles during their
growth. The supersaturation was varied by changing the linear temperature
gradient inside the diffusion chamber.

In a typical growth run at $-15$ C (observing a single needle crystal),
still photos were taken periodically to record the growth after the shutter
was removed. For supersaturations $\sigma _{center}<3.5$ percent surrounding
the growing crystals, ice needles grew slowly into simple columnar
structures. The morphology changed with increasing $\sigma _{center},$ first
to blocky structures on the needle tips, then to thick plates, and then to
thin faceted hexagonal plates at $\sigma _{center}\approx 11$ percent, as
shown in Figure \ref{example}. For $\sigma _{center}>12$ percent, stellar
dendrites and finally fernlike stellar dendrites appeared on the needle
ends, as described in \cite{kgldual14}. The work presented here is limited
to $\sigma _{center}<12$ percent, so the growth morphologies all exhibited
faceted prism surfaces. The growth of these structures could be
quantitatively modeled using a 2D cylindrically symmetric cellular automata
code, as described in \cite{kglca13}, thus avoiding the necessity of a full
3D code.

The radius of the plate (or block) at the end of an ice needle as a function
of time, $R_{plate}(t),$ was extracted directly from the image data, as was
the needle radius, $R_{needle}(t),$ measured at a position 100 $\mu $m below
the needle tip. (We did not distinguish the different \textquotedblleft
radii\textquotedblright\ of a projected hexagonal structure in our image
data, thus limiting the absolute accuracy of our measurements of $R_{plate}$
and $R_{needle}$ to $\pm 5$ percent.) The height of the needle, $H(t),$ was
measured with respect to a \textquotedblleft base
reference\textquotedblright\ that consisted of one or more reference points
in the frost cluster covering the wire substrate at the base of the thin ice
needle (for example, see Figure \ref{example}). The quality and stability of
the base references varied from run to run, and the frost cluster typically
grew and changed with time during a run. As a result, our measurements of $%
H(t)$ were subject to significant uncontrolled systematic errors, making
them less accurate than our measurements of $R_{plate}(t)$ and $%
R_{needle}(t).$

\subsection{Supersaturation in the Diffusion Chamber}

The temperatures of the top and bottom of the second diffusion chamber were
defined by $T_{top,bottom}=T_{center}\pm \Delta T,$ so $T_{top}-T_{bottom}=2%
\Delta T$ (see \cite{kgldual14} for the chamber dimensions). The thermal
characteristics of the chamber walls were designed to produce an accurately
linear temperature gradient within the chamber, as described in \cite%
{kgldual14}. This allows us to use a plane-parallel approximation (moving
the side walls out to infinity) to estimate the water-vapor supersaturation
at the chamber center, where the test crystals were positioned. Solving the
diffusion equation for temperature reproduces the linear temperature profile 
$T(z)$ inside the chamber, with $T_{center}=(T_{top}+T_{bottom})/2.$
Similarly, solving the diffusion equation for water vapor density $c(z)$
also yields a linear function with $c_{center}=(c_{top}+c_{bottom})/2.$ From
these two solutions, the supersaturation at the center of the chamber is%
\begin{eqnarray*}
\sigma _{center} &=&\frac{c_{center}-c_{sat}(T_{center})}{c_{sat}(T_{center})%
} \\
&=&\frac{1}{2}\frac{c_{sat}(T_{top})-2c_{sat}(T_{center})+c_{sat}(T_{bottom})%
}{c_{sat}(T_{center})}
\end{eqnarray*}%
and this expression gives the exact value for $\sigma _{center}$ in the
plane-parallel approximation (ignoring small changes in $D$ with
temperature).

For small $\Delta T$ (where $T_{top}-T_{bottom}=2\Delta T,$ as defined
above), we expand the above expression to obtain%
\begin{eqnarray*}
\sigma _{center} &\approx &\frac{1}{2}\frac{1}{c_{sat}(T_{center})}\frac{%
d^{2}c_{sat}}{dT^{2}}(T_{center})\left( \Delta T\right) ^{2} \\
&\approx &C_{diff}\left( \Delta T\right) ^{2}
\end{eqnarray*}%
and the value of $C_{diff}$ ranges from 0.00282 C$^{-2}$ at -1 C to 0.00314
at -10 C and 0.00332 at -20 C.

\begin{figure}[htb] 
  \centering
  \includegraphics[width=4in,keepaspectratio]{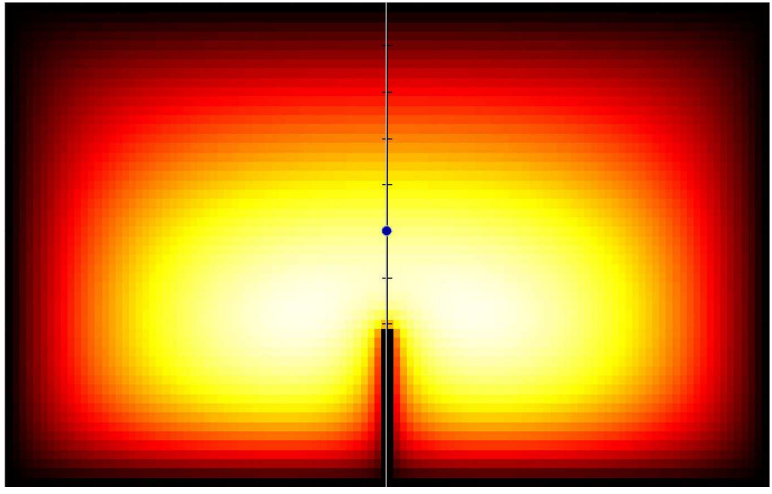}
  \caption{An example numerical model of
the second diffusion chamber, showing a contour plot of the water vapor
supersaturation within the chamber. The top and bottom of the plot are the
top and bottom of the chamber, and the observation point is at the
geometrical center of the chamber, marked here with a round dot. Note that
the supersaturation drops to zero at the chamber walls (dark), and reaches
its maximum value (white) below the center of the chamber. In this
particular model, the supersaturation also goes to zero near an ice-covered
central post that supports the test crystals.}
  \label{sigmamodel}
\end{figure}

We improved upon the plane-parallel approximation by examining a range of
computational models of the diffusion chamber under different conditions,
with an example shown in Figure \ref{sigmamodel}. In these models we solved
the dual-diffusion problem (temperature and water-vapor density) numerically
in three dimensions, performing a number of tests where we changed the
positions of the chamber walls and examined effects of the post supporting
the test crystals. We found that, over a broad range of conditions near $%
T_{center}\approx -15$ C, the side walls reduced $\sigma _{center}$ by a
factor of approximately 0.8 compared to the plane-parallel approximation,
and an ice-covered central stem further reduced $\sigma _{center}$ by a
factor of approximately 0.9. At $T_{center}=-15$ C and small $\Delta T$,
these two factors together changed the supersaturation from the
plane-parallel approximation of $\sigma _{center}\approx 0.0032\left( \Delta
T\right) ^{2}$ to a lower $\sigma _{center}\approx 0.0023\left( \Delta
T\right) ^{2}.$

\begin{figure}[htb] 
  \centering
  \includegraphics[width=4in,keepaspectratio]{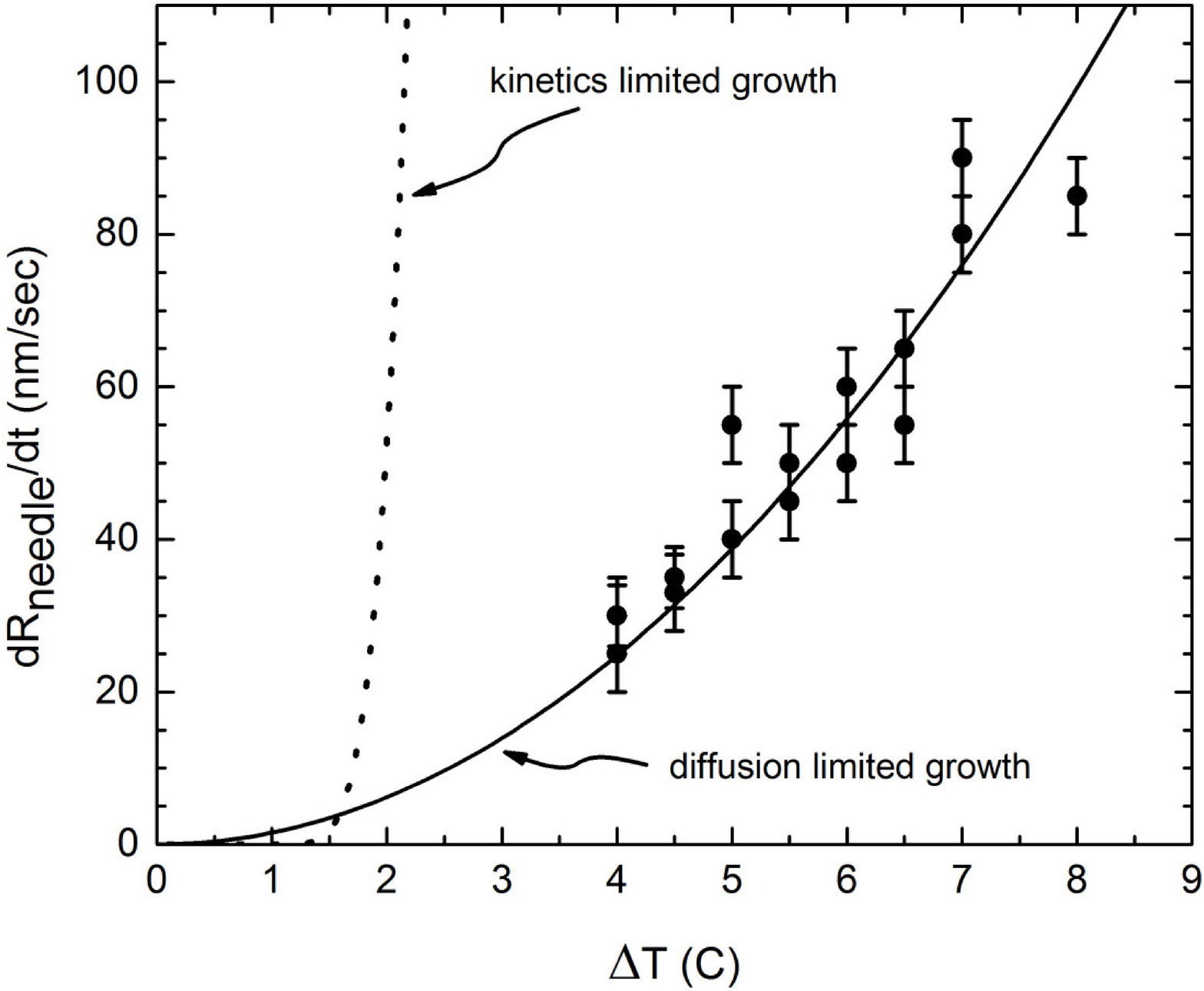}
  \caption{The data points in this graph
show measurements of the needle growth velocity $dR_{needle}/dt$ when $%
R_{needle}=5$ $\mu $m, as a function of the growth-chamber
temperature difference $\Delta T.$ The supersaturation at the chamber center
was proportional to $\Delta T^{2},$ so a purely diffusion-limited growth
model gives $dR_{needle}/dT\sim \Delta T^{2},$ as shown by the solid curve.
In contrast, a model of kinetics-limited growth (dotted curve) indicates
much more rapid growth except at very low $\Delta T$. Thus the data indicate
that the needle growth rate was determined almost entirely by water vapor
diffusion through the surrounding air, independent of $\alpha_{prism}.$ As a result, we found that measurements of $R_{needle}(t)$ served
as an accurate indicator of $\sigma_{center}$ surrounding the test
crystals.}
  \label{drdtat5}
\end{figure}

\section{Quantitative Growth Modeling}

\subsection{Measuring Supersaturation}

In all the measurements presented here, we found that the needle growth $%
R_{needle}(t)$ measured 100 $\mu $m below the needle tip was a good
indicator of the supersaturation surrounding the test crystals. Figure \ref%
{drdtat5} shows $dR_{needle}/dt$ measured at a time (near the beginning of a
growth run) when $R_{needle}=5$ $\mu $m, as a function of the chamber $%
\Delta T$ defined above. A one-parameter quadratic fit yielded $%
dR_{needle}/dt=$1.55$\left( \Delta T\right) ^{2}$ nm/sec, and this curve is
plotted along with the data in Figure \ref{drdtat5}. The dotted line in the
figure shows a purely kinetics-limited growth model with $%
dR_{needle}/dt=\alpha _{prism}v_{kin}\sigma _{center},$ where the kinetic
coefficient of the prism surface is $\alpha _{prism}=\exp (-\sigma
_{0}/\sigma _{center})$ \cite{kglalphas13}, $v_{kin}=208$ $\mu $m/sec \cite%
{libbrechtreview05}, and $\sigma _{center}\approx 0.0023\left( \Delta
T\right) ^{2}$. In this model we took $\sigma _{0}=0.033$ from direct
measurements of kinetics-limited growth \cite{kglalphas13}. The fact that
the dotted curve lies far above the solid curve in Figure \ref{drdtat5}
indicates that the needle growth rate $dR_{needle}/dt$ was primarily
diffusion limited and was therefore independent of $\alpha _{prism}$ to a
good approximation.

We can see that this result is expected by considering the growth of an
infinitely long ice cylinder. Assuming a constant supersaturation $\sigma
_{Rout}$ on a cylindrical outer boundary located at $R_{out},$ an analytic
solution of the diffusion equation gives the growth rate of the ice surface
at $R_{in}$ as 
\begin{eqnarray*}
v(R_{in}) &=&\frac{\alpha _{prism}\alpha _{diffcyl}}{\alpha _{prism}+\alpha
_{diffcyl}}v_{kin}\sigma _{Rout} \\
\alpha _{diffcyl} &=&\frac{1}{B}\frac{X_{0}}{R_{in}}
\end{eqnarray*}%
where $X_{0}=0.145$ $\mu $m (assuming growth at $T=-15$ C in air at a
pressure of one bar) and $B=\log (R_{out}/R_{in})$ \cite{kglca13}. For the
case $\alpha _{diffcyl}\ll \alpha _{prism},$ the growth is limited primarily
by diffusion, giving 
\begin{eqnarray*}
v(R_{in}) &\approx &\alpha _{diffcyl}v_{kin}\sigma _{Rout} \\
&\approx &\frac{X_{0}}{BR_{in}}v_{kin}\sigma _{Rout}
\end{eqnarray*}%
Assuming $\sigma _{Rout}\approx 0.0023\left( \Delta T\right) ^{2}$ from our
chamber modeling calculations above, along with $R_{in}=5$ $\mu $m, we then
obtain%
\[
v(R_{in}=5\mu \rm{m})\approx \frac{13.9}{B}\left( \Delta T\right) ^{2}
\]%
Equating this with the fit curve in Figure \ref{drdtat5} yields $B=8.9,$ or $%
R_{out}\approx 36$ mm. While this is consistent with our expectation for a
distant outer boundary, we cannot say much more because the true outer
boundary in our experiment is more complex than a simple cylindrical
surface. Unlike the spherical case, we cannot simply assume that the outer
boundary is at infinity in the cylindrical case.

This analytical analysis shows that the growth rate $dR_{needle}/dt$ was
mainly diffusion-limited in our measurements, and was thus essentially
independent of $\alpha _{prism}.$ This statement relies on the condition $%
\alpha _{diffcyl}\ll \alpha _{prism}$, which was true for all the
measurements presented here. Thus the diffusion-limited measurements shown
in Figure \ref{drdtat5} are consistent with, and expected from, the
kinetics-limited growth measurements presented in \cite{kglalphas13}.

\section{Detailed Analysis of Test Crystals}

We next look at in-depth analyses of three growth runs at different
supersaturations. These three examples were typical of all our observations,
although we chose the highest quality data sets to analyze further. Because
each test crystal presented somewhat different initial conditions, we found
that analyzing specific test cases was preferable to considering
measurements averaged over many crystals.

\begin{figure}[tb] 
  \centering
  \includegraphics[width=3.8in,keepaspectratio]{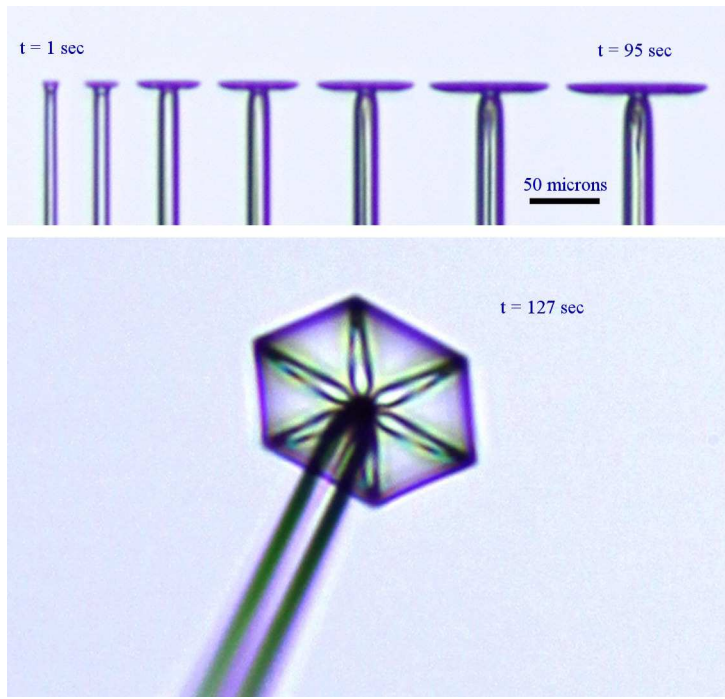}
  \caption{These images show a thin plate growing on the end of an ice needle
in air, when the surrounding supersaturation was $\sigma_{center}\approx 11.3$ percent. The top image shows a composite series of
seven separate images (a subset of the acquired data) taken at different
times, ranging from $t=1$ second (first) to $t=95$ seconds (last). Each
image in the series shows the side view of a faceted, plate-like crystal
growing on the end of a thin ice needle, extracted from images similar to
the one shown in Figure \ref{example}. The lower image shows the
same crystal at $t=127$ seconds, this time with a more frontal view of the
faceted hexagonal plate. Additional observations revealed that the top
surface of the plate was smooth and flat, while six radial, ridge-like
structures formed on the under surface of the plate.}
  \label{highsig}
\end{figure}

\begin{figure}[p] 
  \centering
  \includegraphics[width=3.6in,keepaspectratio]{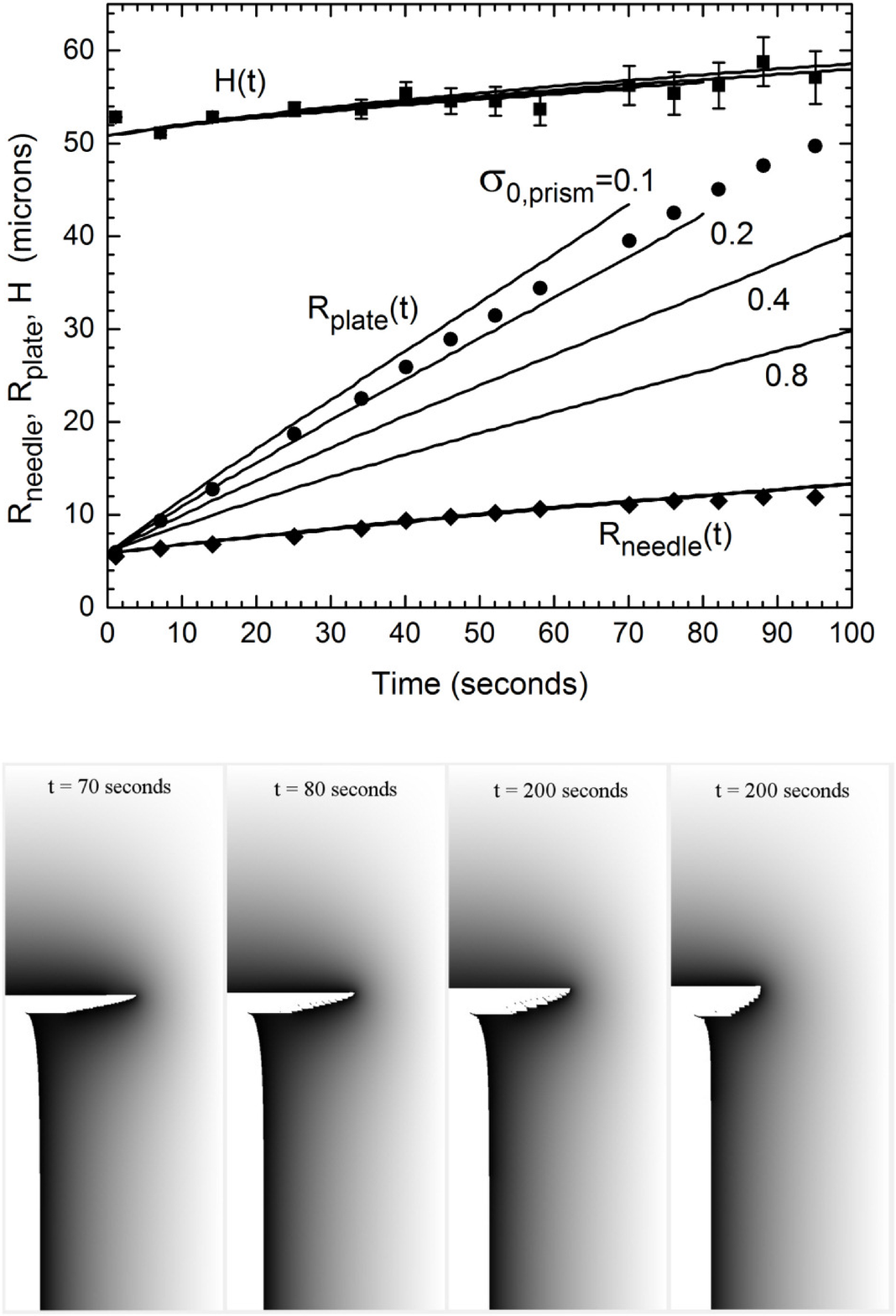}
  \caption{(Top) The three sets of data
points in this graph show measurements of $R_{plate}(t),$ $R_{needle}(t),$
and $H(t)$ (as labeled) extracted from image data including the subset shown
in Figure \protect\ref{highsig}. The measurement of $H(t)$ is shown with an
arbitrary constant subtracted, and the error bars on these points indicate
an estimate of possible systematic errors in the $H\left( t\right) $
measurements. The solid curves in the graph are from four growth models with 
$\protect\sigma _{0,prism}=0.1,$ $0.2,$ $0.4,$ and $0.8$ percent, as
described in the text. The model curves for $R_{needle}(t)$ and $H(t)$ show
little dependence on $\protect\sigma _{0,prism},$ (so the lines overlap) and
the four $R_{plate}(t)$ models are labeled. \protect\break (Bottom) These
four images show ice crystal morphologies corresponding to the four growth
models, with $\protect\sigma _{0,prism}$ increasing from left to right. The
physical times for these model crystals are indicated. The individual images
are in $(r,z)$ space, ranging from $(0,0)$ to the model outer boundary at $%
(r_{\max },z_{\max }).$ The white region in each image shows the ice
crystal, while brightness in the surrounding space is proportional to $%
\protect\sigma /\protect\sigma _{out}.$ For $\protect\sigma _{0,prism}=0.1$
(left crystal), $\protect\alpha _{prism}$ is large and a thin plate grows
rapidly on the end of the ice needle. For $\protect\sigma _{0,prism}=0.8$
(right crystal), the growth model yielded a thicker, slower growing plate.
Note that the models are symmetrical with respect to reflection about the $%
z=0$ plane.}
  \label{highsigdata}
\end{figure}

\subsection{High Supersaturation $\rightarrow $ Thin Plates}

Our first measurement series, shown in Figure \ref{highsig}, shows a
faceted, plate-like crystal growing on the end of an ice needle. It was
taken with $\Delta T=7$ C, at a supersaturation level (equal to the
supersaturation at this position if the crystal were absent) of $\sigma
_{center}\approx 0.0023\left( \Delta T\right) ^{2}\approx 11.3$ percent.
Figure \ref{highsigdata} shows measurements of $R_{plate}(t),$ $R_{needle}(t)
$ at a position 100 $\mu $m below the needle tip, and $H(t)$ extracted from
the image data. Additional measurements with $\Delta T>8$ C yielded
dendritic plate-like structures instead of faceted plates, as described in 
\cite{kgldual14}.

The origin of the time axis was set equal to the time when $R_{plate}\approx
R_{needle},$ so the crystal morphology was that of a simple column at $t=0.$
The $t=0$ point was typically determined from an extrapolation of $t>0$ data.

We modeled these data using the cylindrically symmetric cellular automata
model described in \cite{kglca13}, including the modification for surface
diffusion described in \cite{kglSDmodel}. The results of several model
calculations are shown together with the experimental data in Figure \ref%
{highsigdata}. Typical model parameters included a boundary box with $%
r_{\max }=500$ pixels = 72.5 $\mu $m and $z_{\max }=1250$ pixels = 181 $\mu $%
m, $z_{init}=100$ $\mu $m (the initial needle height), $T=-15$ C, $%
N_{speed}=100$ \cite{kglSDmodel}, Gibbs-Thomson parameter $\delta =0.3$ nm 
\cite{kglca13}, and surface diffusion lengths $x_{s}=14$ nm for both facets 
\cite{furukawa14, kglSDmodel, kglSD}. The models yielded $R_{plate}(t),$ $%
R_{needle}(t),$ and $H(t)$ for comparison with data, plus the crystal
morphology and supersaturation near the crystal surface.

\begin{figure}[tb] 
  \centering
  \includegraphics[width=3.7in,keepaspectratio]{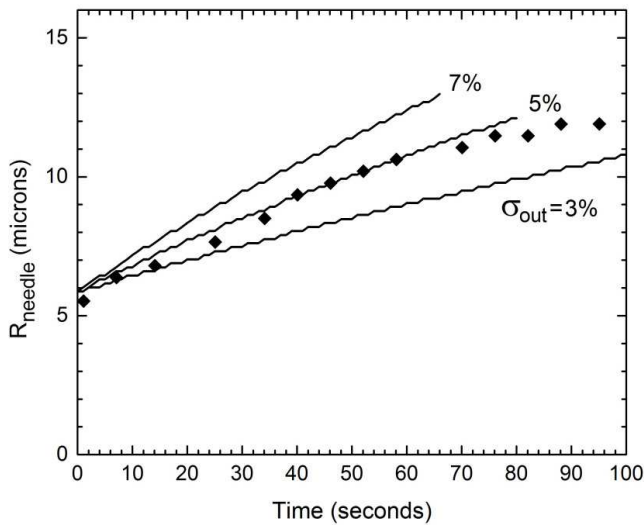}
  \caption{This graph shows the same $%
R_{needle}(t)$ data points as in Figure \protect\ref{highsigdata}, along
with several model calculations using different values of $\protect\sigma %
_{out}.$ Over a broad range of measurement conditions, our model
calculations verified that $R_{needle}(t)$ was sensitive to $\protect\sigma %
_{out}$ while being relatively insensitive to other model parameters.}
  \label{Rneedle}
\end{figure}

We first used the numerical models to verify that $R_{needle}(t)$ was
sensitive to $\sigma _{out}$ (the supersaturation at the outer boundary of
the model box) and that $R_{needle}(t)$ was insensitive to other model
parameters, including $\alpha _{prism}.$ Figure \ref{Rneedle} shows the same 
$R_{needle}(t)$ data as in Figure \ref{highsigdata} along with several model
crystals. As seen in the figure, fitting the data required a model
supersaturation of $\sigma _{out}\approx 5$ percent. As expected from the
above discussion, additional models (not shown) confirmed that $%
R_{needle}(t) $ was insensitive to other model parameters, including $\alpha
_{prism}.$

The fact that the best fit $\sigma _{out}\approx 5$ percent is smaller than
our estimated $\sigma _{center}\approx 11.3$ percent arises from the finite
size of the model box, which was much smaller than any reasonable estimate
for the outer boundary of the experimental system (for example the 36 mm
described above). Using $R_{out}=72.5$ $\mu $m and $R_{in}=5$ $\mu $m gives $%
B=\log (R_{out}/R_{in})=2.7,$ which is smaller than the $B$ value above by a
factor of 0.3, comparable to $5/11.3=0.44.$ In other words, $\sigma _{out}$
is smaller than $\sigma _{center}$ because $\sigma _{center}$ is the
supersaturation far from the growing crystal, while $\sigma _{out}$ gives
the supersaturation quite close to the crystal. If the outer dimensions of
the model space were orders of magnitude larger, we expect that the best fit 
$\sigma _{out}$ would be much closer to $\sigma _{center}.$ To compensate
for the small model size, we chose $\sigma _{out}$ to fit the experimental
data.

What this meant in practice is that we could use the $R_{needle}(t)$ data to
accurately determine the correct $\sigma _{out}$ in the models, because $%
R_{needle}(t)$ is quite sensitive to this parameter, while being insensitive
to other model parameters. Therefore, for the discussion that follows, we
adjusted $\sigma _{out}$ to fit the $R_{needle}(t)$ data, and then did not
change $\sigma _{out}$ further. Because the needle growth was mainly limited
by diffusion, the needle surface acted as a \textquotedblleft witness
surface\textquotedblright\ to accurately constrain $\sigma _{out}$ in our
models. We found this to be true over a broad range of experimental
conditions, and this ended up being a quite beneficial feature of growing
ice crystals on thin ice needles using our dual-chamber apparatus, as it
allowed better quantitative analysis than would have been possible otherwise.

Having used $R_{needle}(t)$ to fix $\sigma _{out}=5$ percent, we generated
the four growth models in Figure \ref{highsigdata} using \cite{kglalphas13} $%
\alpha _{basal}=A_{basal}\exp (-\sigma _{0,basal}/\sigma _{surface})$ with $%
[A_{basal},\sigma _{0,basal}]=[1,2]$ (displaying $\sigma _{0,basal}$ here in
percent), and similarly $[A_{prism},\sigma _{0,prism}]=[1,x]$ with $%
x=(0.1,0.2,0.4,0.8)$ percent and $R_{init}=6$ $\mu $m. Note that the four
curves for $R_{needle}(t)$ lie on top of one another in Figure \ref%
{highsigdata}, confirming that $R_{needle}(t)$ depends mainly on $\sigma
_{out},$ which was the same for these four models.

The values for $A_{prism},$ $A_{basal},$ and $\sigma _{0,basal}$ were taken
from the measurements in \cite{kglalphas13}, as was the $\alpha $
parameterization for both facets. (This parameterization describes growth
that is limited by 2D nucleation on atomically flat surfaces \cite%
{libbrechtreview05}.) The data show, however, that $\sigma _{0,prism}$
needed to be near 0.15 percent, which is much smaller than the $\sigma
_{0,prism}=3.3$ percent reported in \cite{kglalphas13}. These results
quantify, for a specific parameterization of $\alpha \left( \sigma \right) $%
, that we must have $\alpha _{prism}\gg \alpha _{basal}$ to form a thin ice
plate, as one would expect. The results also indicate $\alpha
_{prism}\approx 1$ for the edge of the plate-like crystal, which is
consistent with the observation that the plate growth is no longer faceted
if $\Delta T$ is slightly larger. The utility of these results becomes
apparent when we compare different measurements in an attempt to form a
global growth model, as we discuss below, after we first examine additional
experimental data.

\begin{figure}[htb] 
  \centering
  \includegraphics[width=3.9in,keepaspectratio]{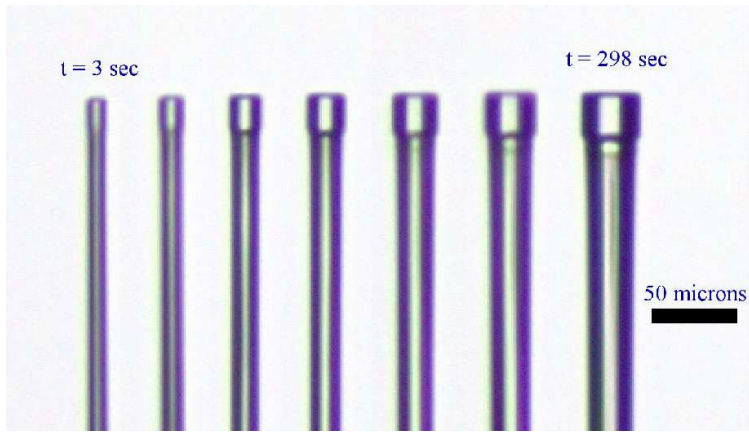}
  \caption{This composite image shows a
faceted ice block growing on the end of an ice needle, when the surrounding
supersaturation was $\sigma_{center}\approx 4.6$ percent.
Measurements of the growing crystal extracted from these images are shown in
Figure \ref{lowsigdata}. Note that the full data set included more
images than are shown here.}
  \label{lowsig}
\end{figure}

\begin{figure}[p] 
  \centering
  \includegraphics[width=3.45in,keepaspectratio]{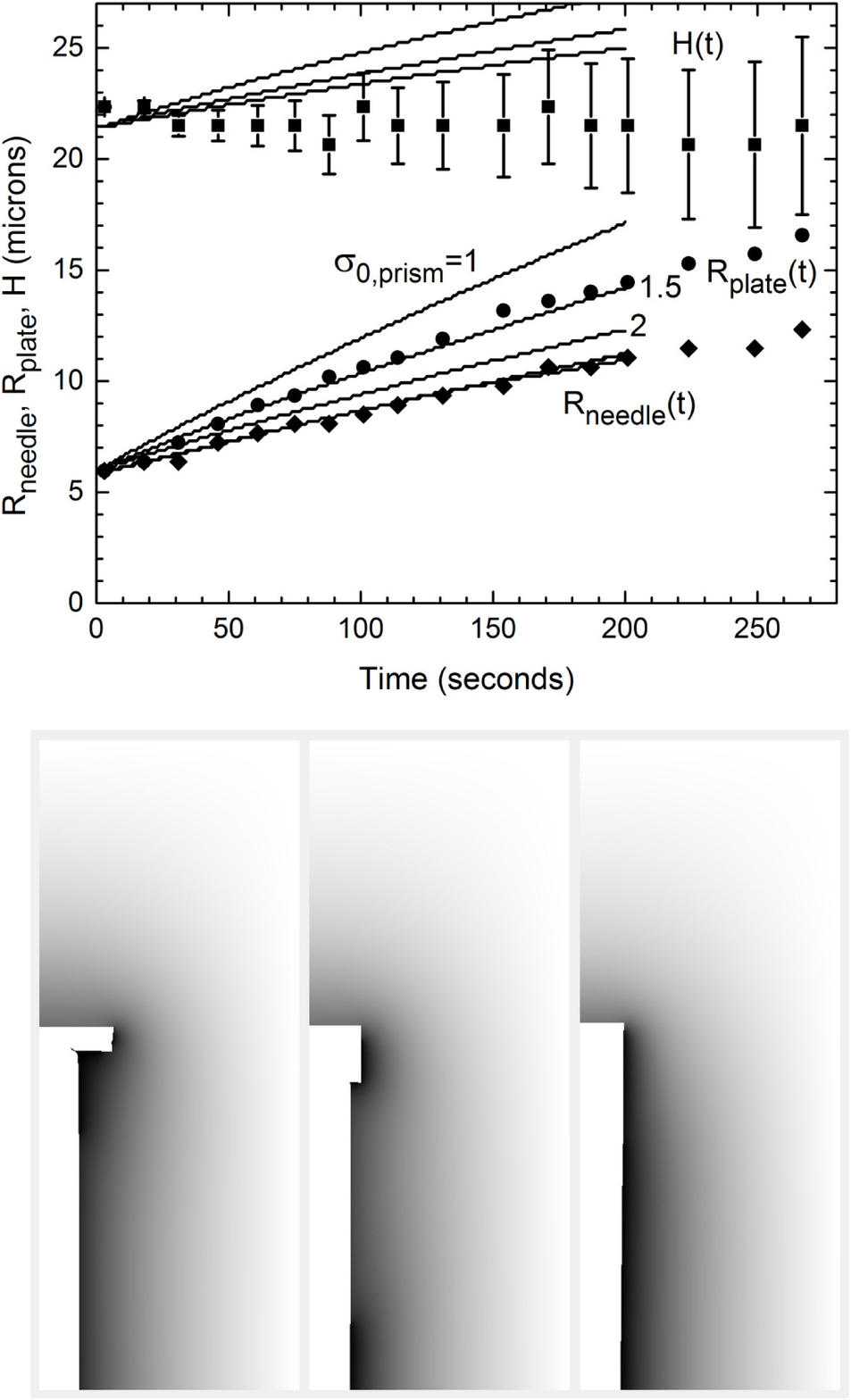}
  \caption{(Top) Similar to Figure \protect
\ref{highsigdata}, the three sets of data points in this graph show
measurements of $R_{plate}(t),$ $R_{needle}(t),$ and $H(t)$ extracted from
image data including the subset shown in Figure \protect\ref{lowsig}. Again
the measurements of $H(t)$ are shown with an arbitrary constant subtracted,
and the error bars on these points indicate an estimate of possible
systematic errors in the $H\left( t\right) $ measurements. The solid curves
in the graph are from three growth models with $\protect\sigma _{0,prism}=1,$
$1.5,$ and $2$ percent, as described in the text. Again the (overlapping)
model curves for $R_{needle}(t)$ showed little dependence on $\protect\sigma %
_{0,prism},$ and the three $R_{plate}(t)$ models are labeled. \protect\break %
(Bottom) Also similar to Figure \protect\ref{highsigdata}, these three plots
show ice crystal growth morphologies corresponding to the three growth
models in the top graph, with $\protect\sigma _{0,prism}$ increasing from
left to right. The physical times are at $t=200$ seconds for all three model
crystals. Note the transition from a simple columnar structure at $\protect%
\sigma _{0,prism}=2$ percent (right crystal) to the formation of a thick
plate on the needle end at $\protect\sigma _{0,prism}=1$ percent (left
crystal).}
  \label{lowsigdata}
\end{figure}

\subsection{Low Supersaturation $\rightarrow $ Blocky Columns}

Our next set of data was taken at $\Delta T=4.5$ C, so the supersaturation
surrounding the growing crystal was $\sigma _{center}\approx 0.0023\left(
\Delta T\right) ^{2}\approx 4.6$ percent. Several example images are shown
in Figure \ref{lowsig}. In contrast to the high-supersaturation case
described in the previous section, here we see a faceted block growing on
the end of the ice needle. Additional measurements with $\Delta T<4$ C
yielded basic columnar structures with some negative tapering (the tip wider
than the base).

We modeled these data using the model parameters described above, except our
best fit to $R_{needle}(t)$ was obtained with $\sigma _{out}=1.7$ percent.
Again we used $[A_{basal},\sigma _{0,basal}]=[1,2]$ and $[A_{prism},\sigma
_{0,prism}]=[1,x]$ with $x=(1,1.5,2)$ and $R_{init}=6$ $\mu $m. The results
of these models are shown in Figure \ref{lowsigdata} along with the
corresponding crystal morphologies.

We obtained a good fit to the data with $\sigma _{0,prism}=1.5$ percent,
which is much higher than the high-sigma value of $\approx 0.15$ percent.
And again, the $\sigma _{0,prism}=1.5$ percent model fits both the
morphology and quantitative growth behavior of the crystal. Note that the
parameterization $\alpha _{prism}=A_{prism}\exp (-\sigma _{0,prism}/\sigma
_{surface})$ includes an intrinsic supersaturation dependence, and the usual
assumption for nucleation-limited growth is that $\sigma _{0,prism}$ is
constant \cite{kglalphas13}. The meaning of a changing $\sigma _{0,prism}$
value is discussed below.

The models did not accurately reproduce the $H(t)$ data in Figure \ref%
{lowsigdata}, but we believe that the substantial systematic errors in our $%
H(t)$ measurements (described above) could explain the discrepancy. The
experimental uncertainties in our $H(t)$ data are great enough that we could
not obtain much useful information about $\sigma_{0,basal}$ using needle
growth measurements. The change in $\sigma_{0,prism}$, by contrast, is a
robust result from these data.

\begin{figure}[htb] 
  \centering
  \includegraphics[width=4.2in,keepaspectratio]{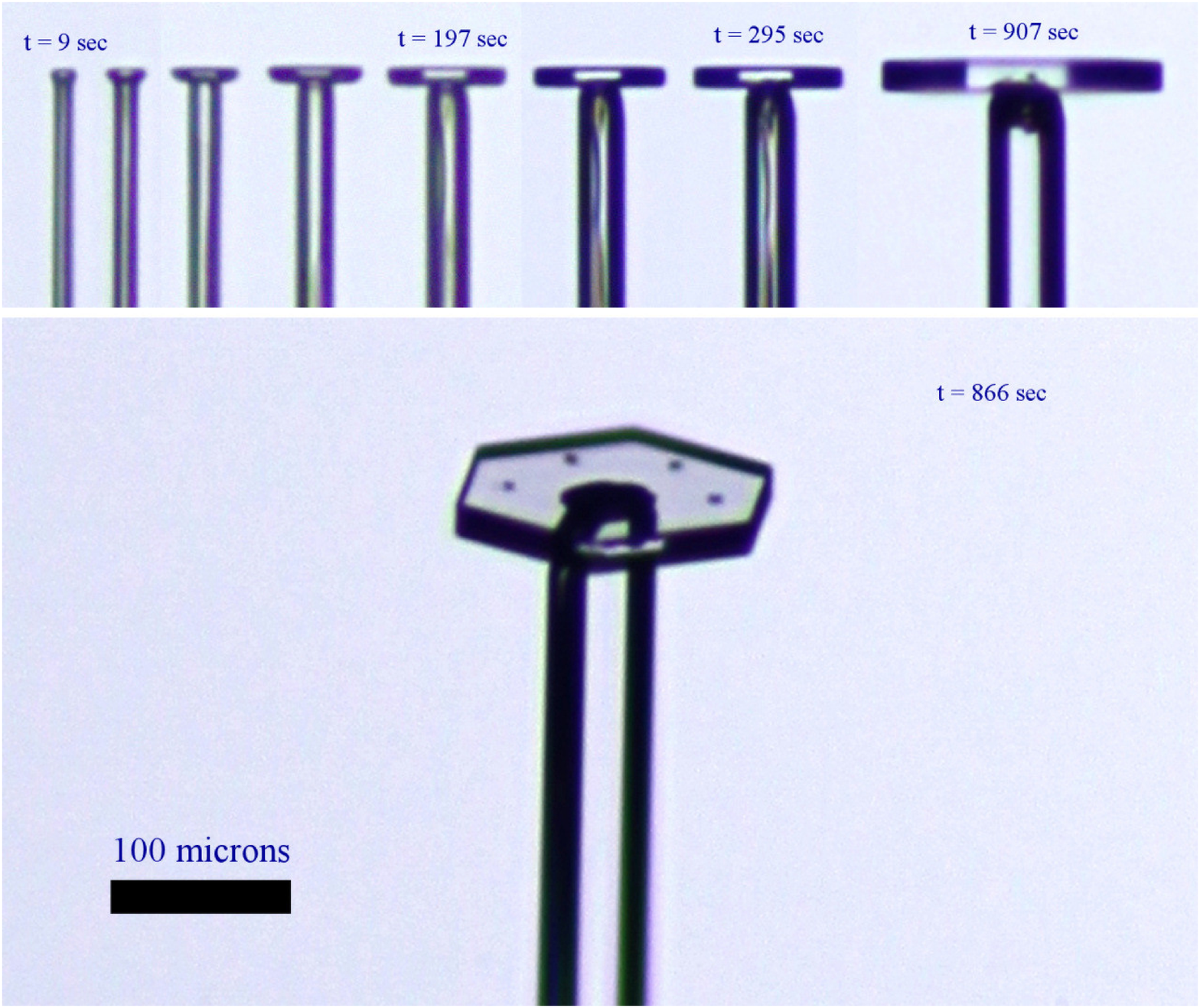}
  \caption{Another composite image showing a thick, faceted ice plate growing on the
end of an ice needle, when the surrounding supersaturation was $\sigma_{center}\approx 7.0$ percent. Measurements of the growing crystal
extracted from these images are shown in Figure \ref{lowsigdata}.
The full data set included more images than are shown here.}
  \label{intsigma}
\end{figure}

\begin{figure}[p] 
  \centering
  \includegraphics[width=4in,keepaspectratio]{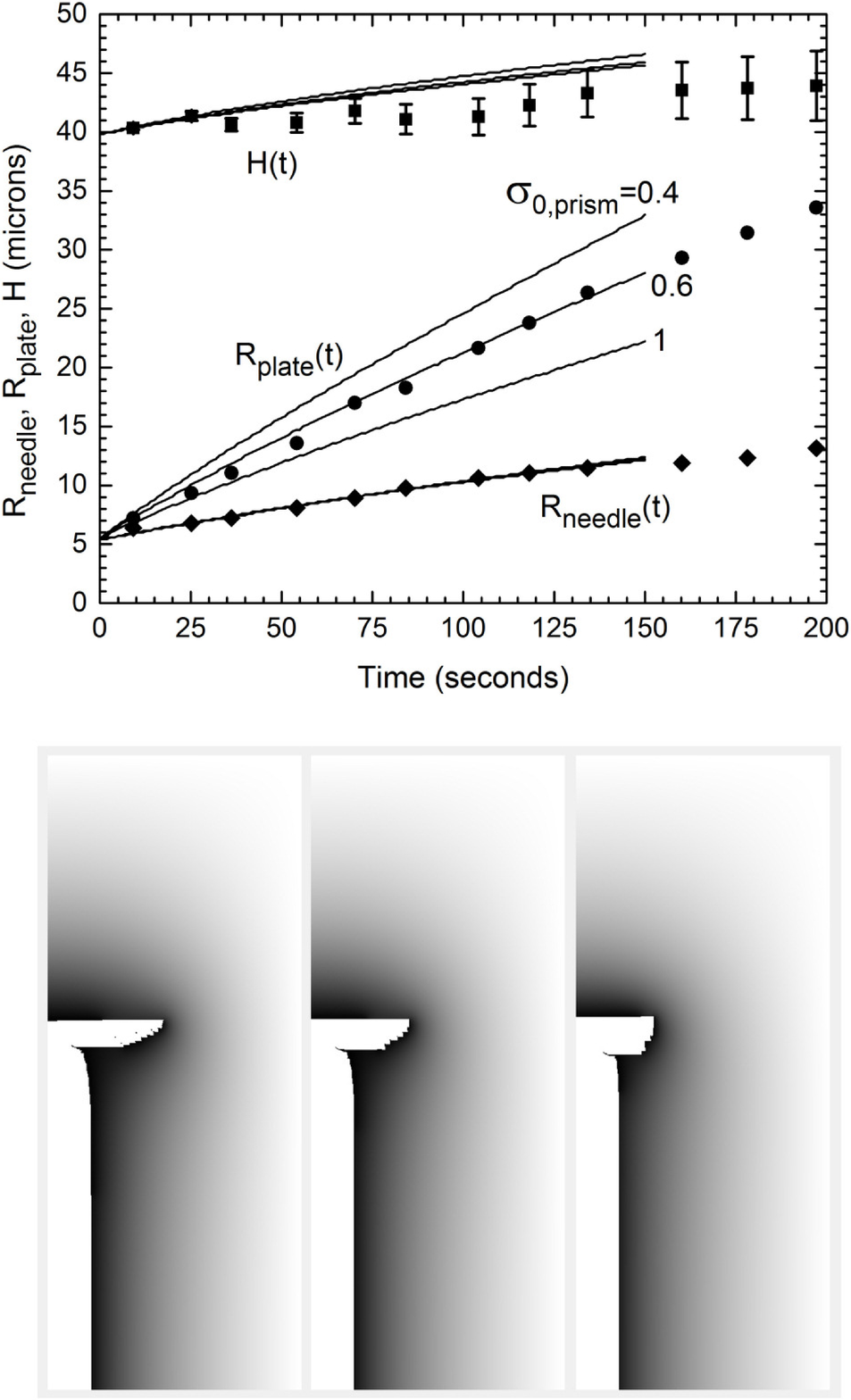}
  \caption{Similar to the previous two
examples, this shows a comparison of measurements and growth models of the
formation of a thick plate on the end of an ice needle at $\protect\sigma %
_{center}\approx 7.0$ percent, as described in the text. The morphology
images are at $t=150$ seconds.}
  \label{intersigdata}
\end{figure}

\subsection{Intermediate Supersaturation $\rightarrow $ Thick Plates}

Finally, Figure \ref{intsigma} shows a run taken at $\Delta T=5.5$ C, with
the corresponding supersaturation around the crystals $\sigma
_{center}\approx 0.0023\left( \Delta T\right) ^{2}\approx 7.0$ percent. At
this intermediate supersaturation we see a thick, faceted plate growing on
the end of the ice needle, intermediate between the preceding high- and low-$%
\sigma _{center}$ data sets, as one would expect. Measurements from this
data set are shown in Figure \ref{intersigdata} along with several growth
models.

The three models shown in Figure \ref{intersigdata} used $[A_{basal},\sigma
_{0,basal}]=[1,2]$ and $[A_{prism},\sigma _{0,prism}]=[1,x]$ with $%
x=(0.4,0.6,1)$, along with model parameters $\sigma _{out}=3$ percent and $%
R_{init}=5.6$ $\mu $m. As with the previous two example crystals, we found
we could fit the crystal morphology and quantitative growth rates reasonably
well with a fairly simple model, mainly adjusting $\sigma _{out}$ to fit the 
$R_{needle}(t)$ data and then adjusting $\sigma _{0,prism}$ to fit the $%
R_{plate}(t)$ data. In all cases the model behavior was relatively
insensitive to the other input parameters, provided these were sensibly
chosen.

The model calculations showed that the plate thickness increased with time
via growth of the top surface of the plate, with very little growth of the
lower surface. This provided some insight into systematic errors in $H(t),$
since we could use the plate thickness as an additional measurement of $H(t)$
that did not include a far-away base reference. We found that the new $H(t)$
data fit the model calculations quite well using $\sigma _{0,basal}=2,$
while the $H(t)$ data shown in Figure \ref{intersigdata} did not match the
models as well. This again suggested that the data presented here are likely
consistent with the intrinsic value of $\sigma _{0,basal}=2$ percent
presented in \cite{kglalphas13}. The relatively poor model fits to $H(t)$
may arise entirely from systematic errors in our measurements of $H(t)$, as
described above. Put another way, the $H(t)$ measurements from the
dual-chamber data presented here are not accurate enough to yield solid
conclusions about $\sigma _{0,basal}.$ In contrast, the $R_{plate}(t)$ and $%
R_{needle}(t)$ data are quite accurate, yielding robust conclusions from a
careful examination of the data and growth models.

\subsection{Model Comparisons}

It is instructive to use the growth models to extract information about the
supersaturation at the surface of the growing test crystals. For example, a
model of the thin-plate crystal with $\sigma _{0,prism}=0.15$ percent (which
fits the data quite well, as seen in Figure \ref{highsigdata}) indicates
that $\sigma _{prism-surface}\approx 0.50\pm 0.05$ percent on the edge of
the plate, a value that is nearly constant in time (yielding the nearly
constant $dR_{plate}/dt$ seen in Figure \ref{highsigdata}). Similarly, the
model of the blocky crystal with $\sigma _{0,prism}=1.5$ percent (shown in
Figure \ref{lowsigdata}) gives $\sigma _{prism-surface}\approx 0.47\pm 0.07$
percent on the top edge of the block. Thus we see that the supersaturation
at the prism surface of the fast-growing plate is not substantially higher
than the supersaturation at the prism surface of the slow-growing block.

Changing the parameterization of $\alpha _{prism}$ does not significantly
change this somewhat counter-intuitive result. Switching to models with a
constant $\alpha _{prism}$ (independent of supersaturation), we find that a
high $\alpha _{prism}\approx 1$ is needed to reproduce the thin-plate data
(both the morphology and growth rates), while a much lower $\alpha
_{prism}\approx 0.04$ is necessary to reproduce the blocky-crystal data. For
both these $\alpha _{prism}$ parameterizations, we find that a large change
in $\alpha _{prism}$ is responsible for the large change in morphology and
growth rates; the supersaturation at the prism surface actually changes very
little between these two cases.

This result strongly supports the hypothesis \cite{sdak12} that $\alpha
_{prism}$ is a strong function of the width $w_{prism}$ of the prism facet
when $w_{prism}$ approaches atomic dimensions, at least for ice growth from
water vapor near -15 C. Other hypotheses are excluded by observations
indicating that: 1) on a broad prism facet, the functional form $\alpha
_{prism}=\exp (-\sigma _{0,prism}/\sigma _{prism-surface})$ (with constant $%
\sigma _{0,prism}$) fits low-pressure growth data over a broad range of
growth conditions \cite{sdak12, kglalphas13}, and 2) $\sigma _{prism-surface}
$ is about the same for the thin-plate and blocky column examples shown
above. If we reject the hypothesis that $\alpha _{prism}$ depends on the
width of the prism facet, there remains (in our opinion) no physically
plausible way to explain all the data.

\begin{figure}[htb] 
  \centering
  \includegraphics[width=3.7in,keepaspectratio]{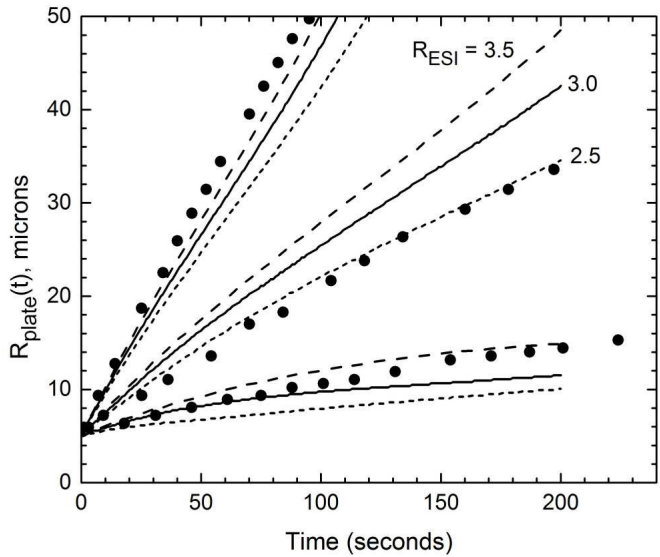}
  \caption{This plot compares our
comprehensive model of ice crystal growth from water vapor at -15 C with the
three data sets described above, plotting $R_{plate}(t)$ data for the three
cases (points). The three sets of lines show models with $R_{ESI}=2.5$
(dotted), 3.0 (solid), and 3.5 (dashed) microns. These models were each run
with $\protect\sigma _{out}=1.7,$ $3,$ and $5$ percent for comparison with
the three sets of data. Although we were not able to match the three sets of
data precisely using our simple model of $\protect\sigma _{0,prism}$, the
comprehensive model does reproduce the transition from blocky column (lower
data points) to thin plate (upper data points) with reasonable accuracy. We
see that the underlying cause of the transition is an edge-sharpening
instability\ (ESI) that causes this rather abrupt change in growth
morphology as $\protect\sigma _{out}$ is increased.}
  \label{comprehensive1}
\end{figure}

\begin{figure}[htb] 
  \centering
  \includegraphics[width=3.7in,keepaspectratio]{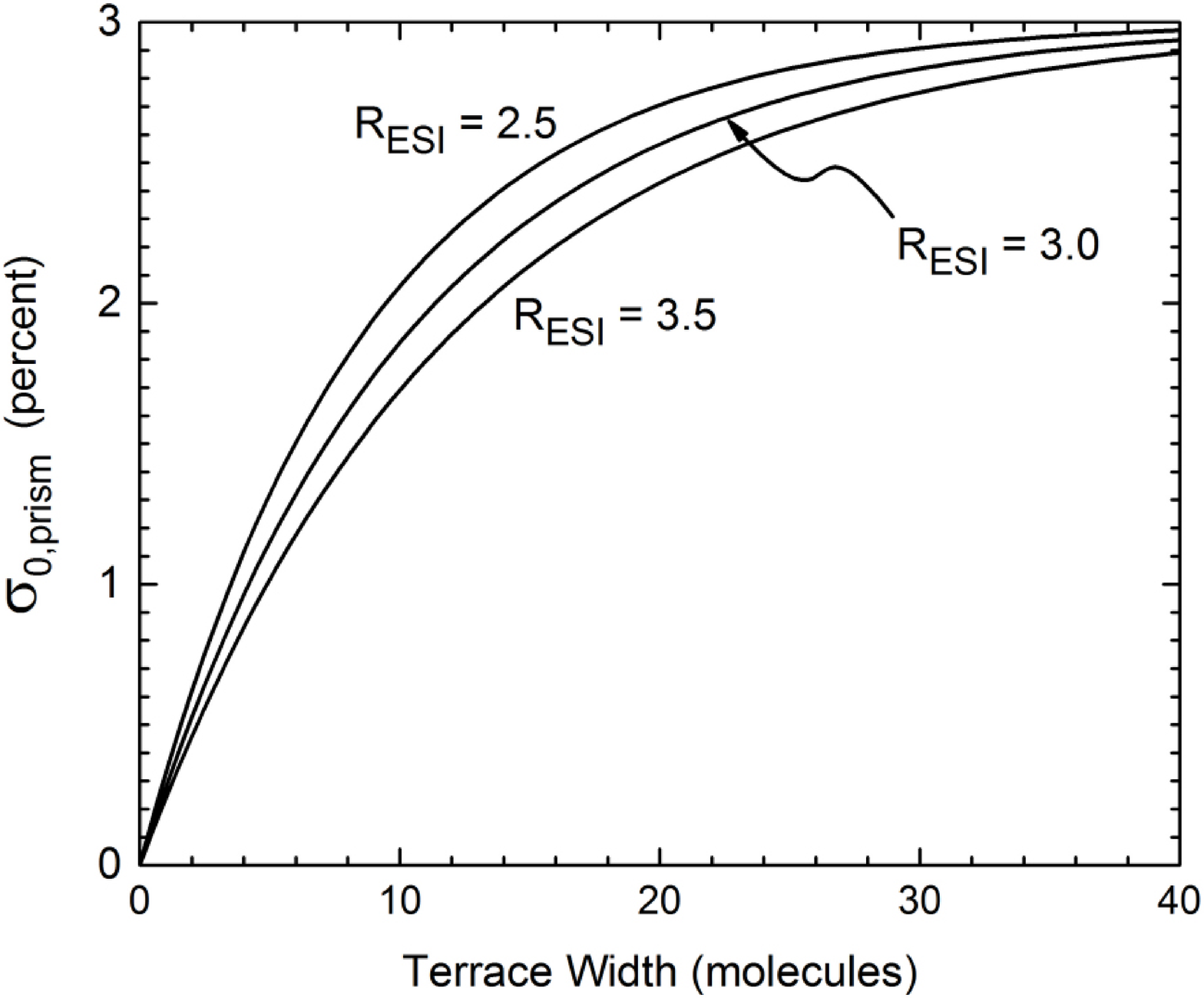}
  \caption{This plot shows $\protect\sigma %
_{0,prism}$ as a function of the width of the uppermost prism terrace, for
the growth model described in the text, with different values of the model
parmeter $R_{ESI}$ (in microns).}
  \label{comprehensive2}
\end{figure}

\section{A Comprehensive Ice Growth Model}

Allowing $\alpha _{prism}$ to vary with the width of the prism facet, we
found it straightforward to devise a single ice growth model that reproduced
all the data presented above. We added a single new parameter to our above
parameterization of $\alpha _{prism}$, changing it to $\alpha _{prism}=\exp
(-\sigma _{0,prism}/\sigma _{prism-surface})$ with $\sigma _{0,prism}=\sigma
_{0,prism,\infty }(1-\exp (-R_{c}/R_{ESI})),$ where $R_{ESI}$ is the new
model parameter and $R_{c}=$ $2\Delta rN_{z}$ is a surrogate for the width
of the prism facet (here $N_{z}$ is the number of model pixels comprising
the outermost prism facet, as described in \cite{kglca13}). We assumed $%
\sigma _{0,prism,\infty }=3$ percent from the measurements in \cite%
{kglalphas13}.

Applying this new model to the three data sets described above yields the
result shown in Figure \ref{comprehensive1}. These models use $\sigma
_{out}=1.7,$ $3,$ and $5$ percent, $R_{init}=5$ $\mu $m, $[A_{basal},\sigma
_{0,basal}]=[1,2]$ and $A_{prism}=1.$ As seen in the plot, adding the
additional $R_{ESI}$ model parameter allows us to reproduce the
morphological transition from blocky column to thin plate with increasing
supersaturation. The morphologies of the models were similar to the best-fit
constant-$\sigma _{0,prism}$ models shown above.

Figure \ref{comprehensive2} shows how $\sigma _{0,prism}$ depends on the
width $w_{prism}$ of the uppermost prism terrace in our model. Here we have
converted the model parameter $R_{c}$ to $w_{prism}$ using $%
w_{prism}=N_{z}a, $ where $a\approx 0.3$ nm is the size of a water molecule.
(A discussion of this conversion is described in detail in \cite{kglSDmodel}%
). This plot shows that $\sigma _{0,prism}$ is essentially equal to $\sigma
_{0,prism,\infty }$ when the terrace width is greater than $\sim 30$
molecules. And for such broad terraces, we use the measured $\sigma
_{0,prism,\infty }$ from \cite{kglalphas13}. But when $w_{prism}$ becomes
smaller, $\sigma _{0,prism}$ decreases until $\sigma _{0,prism}\approx 0$
(and therefore $\alpha _{prism}\approx 1)$ when $w_{prism}\rightarrow 0.$

We can only speculate as to why $\sigma _{0,prism}(w_{prism})$ has the form
shown in Figure \ref{comprehensive2}; we have no detailed molecular model
that would explain this behavior. However, there is also no molecular model
at present that can explain the measured $\sigma _{0,prism}(T)$ for large
prism facets \cite{kglalphas13}. The latter derives from the change in the
step energy of a prism facet with temperature \cite{kglalphas13}, but step
energies in ice have received little experimental and theoretical attention
to date. We believe that molecular dynamics simulations could shed
considerable light on why the step energies in ice behave as they do, but
for now we do not understand this rather fundamental aspect of ice
energetics.

\section{Discussion}

In summary, we have described a comprehensive model of ice growth from water
vapor at -15 C that reproduces both morphologies and measured growth rates
with reasonable fidelity over a range of supersaturations. In particular,
our model nicely explains the observed morphological transition from blocky
columns to thin plates growing on the ends of ice needle crystals. This
transition is described as an edge-sharpening instability in the ice growth
behavior, brought about by a sharp reduction in the nucleation barrier on a
prism facet (parameterized by $\sigma _{0,prism}$) when the width of the
facet approaches molecular dimensions.

Our quantitative computer modeling of carefully controlled growth
measurements drove the development of this picture of ice growth dynamics.
By comparing the above measurements with those described in \cite%
{kglalphas13}, we found no viable alternative to the hypothesis that $\alpha
_{prism}$ changes with the width of the prism terrace. Since the 2D
nucleation model (defining $\sigma _{0,prism}$ from the molecular step
energy of the ice surface via classical nucleation theory) fits the ice
growth so well in \cite{kglalphas13}, it was a modest conceptual step to
model the change in $\alpha _{prism}$ as a change in $\sigma _{0,prism},$ as
we did above.

To our knowledge, this is the first time a kinetic model of the
diffusion-limited growth of ice has successfully reproduced measured growth
behaviors over a range of supersaturations. We believe this is a significant
step toward out ultimate goal of modeling ice growth behavior as a function
of both temperature and supersaturation. The model parameters determined
from these efforts (such as the molecular step energies on faceted ice
surfaces) should lead to a better understanding of the surface structure and
dynamics of ice crystals and ice crystallization.

We acknowledgement support from the Cambridge-Caltech Exchange Program,
Caltech's WAVE Fellows Program, and the SURF program at Caltech.

\bibliography{C:/Dropbox/1-kgl-top/Papers/1Bibliography/kglbiblio3}
\bibliographystyle{unsrt}  

\end{document}